\documentclass[final]{aipproc}
\layoutstyle{6x9}

\usepackage{amssymb,amsmath}

\begin{document}

\title{Phenomenology of CP Violation \\ in a Flavor Blind MSSM and Beyond}


\author{Wolfgang Altmannshofer}{
address={Physik Department, Technische Universit\"at M\"unchen, D-85748 Garching, Germany}
}

\begin{abstract}

We present an analysis of low energy CP violating observables in the Minimal Supersymmetric Standard Model (MSSM). We focus on the predictions of CP violation in $b \to s$ transitions in the framework of a flavor blind MSSM, where the CKM matrix remains the only source of flavor violation, but additional CP violating phases are introduced in the soft SUSY breaking sector.
We find large and strongly correlated effects in $\Delta F=0$ observables like the electric dipole moments (EDMs) of the electron and the neutron, as well as in $\Delta F=1$ observables like the time dependent CP asymmetries in $B \to \phi K_s$ and $B \to \eta^\prime K_s$, the direct CP asymmetry in $b \to s \gamma$ and in several CP asymmetries in $B \to K^* \mu^+ \mu^-$.
On the other hand, observables that are only sensitive to CP violation in $\Delta F=2$ transitions, in particular the $B_s$ mixing phase, are found to be SM like in this framework.
We stress that only in presence of additional sources of {\it flavor violation}, sizeable New Physics effects to {\it CP violation} in meson mixing can occur.

\end{abstract}

\maketitle

\section{Hints for New Sources of CP Violation}

Although the Standard Model (SM) CKM picture of flavor and CP violation has been confirmed over the last years at the level of (10-20)\%~\cite{Barberio:2008fa}, there are in fact hints of discrepancies with respect to some SM expectations (see e.g.~\cite{Lunghi:2008aa}):
\begin{itemize}
\item [i)] the measured amount of CP violation in $B_d$ mixing ($S_{\psi K_S}$) seems insufficient to explain CP violation in $K$ mixing ($\epsilon_K$);
\item [ii)] the time-dependent CP asymmetries in the loop induced decays $B \to \phi K_S$ and $B \to \eta^\prime K_S$ ($S_{\phi K_S}$ and $S_{\eta^\prime K_S}$) are measured to be considerably smaller than $S_{\psi K_S}$;
\item [iii)] recent analyses find a $B_s$ mixing phase much larger than the tiny SM prediction. 
\end{itemize}
Taking these tensions seriously, a natural way to address them is to go beyond the SM and to introduce new CP violating phases.

\section{Phenomenology of CP Violation in the MSSM}

The MSSM contains many free parameters that can provide additional sources of CP violation. Once (some of) these parameters are assumed to be complex, in general several CP violating processes will receive NP contributions simultaneously. In the following we consider observables that are sensitive to CP violation in
\begin{itemize}
\item $\Delta F=0$ amplitudes, like the EDMs of the electron and neutron, $d_e$ and $d_n$;
\item $\Delta F=1$ amplitudes, like the direct CP asymmetry in the $b \to s \gamma$ decay, $A_{CP}(b\to s\gamma)$ or the CP asymmetries in the $B \to K^* \mu^+ \mu^-$ decay;
\item $\Delta F=2$ amplitudes, like $\epsilon_K$, that measures the amount of CP violation in $K$ mixing or the mixing induced CP asymmetries in $B \to \psi K_S$ and $B_s \to \psi \phi$, $S_{\psi K_S}$ and $S_{\psi \phi}$, that measure the $B_d$ and $B_s$ mixing phases;
\item both $\Delta F=1$ and $\Delta F=2$ amplitudes, like the time dependent CP asymmetries in $B \to \phi K_S$ and $B \to \eta^\prime K_S$, $S_{\phi K_S}$ and $S_{\eta^\prime K_S}$.
\end{itemize}

\subsection{A Flavor Blind MSSM with CP Violating Phases}

In the following we focus on the phenomenology of a so called Flavor Blind MSSM, a rather restricted framework where the CKM matrix remains the only source of flavor violation, but additional CP violating phases are introduced in the soft sector. In particular, in~\cite{ABP} we assumed flavor universal squark masses, flavor diagonal but hierarchical trilinear couplings, but allowed the trilinear couplings to be complex.\footnote{For analyses of similar frameworks see~\cite{Baek:1998yn}.}

\begin{figure}[t]
\includegraphics[width=.29\textwidth]{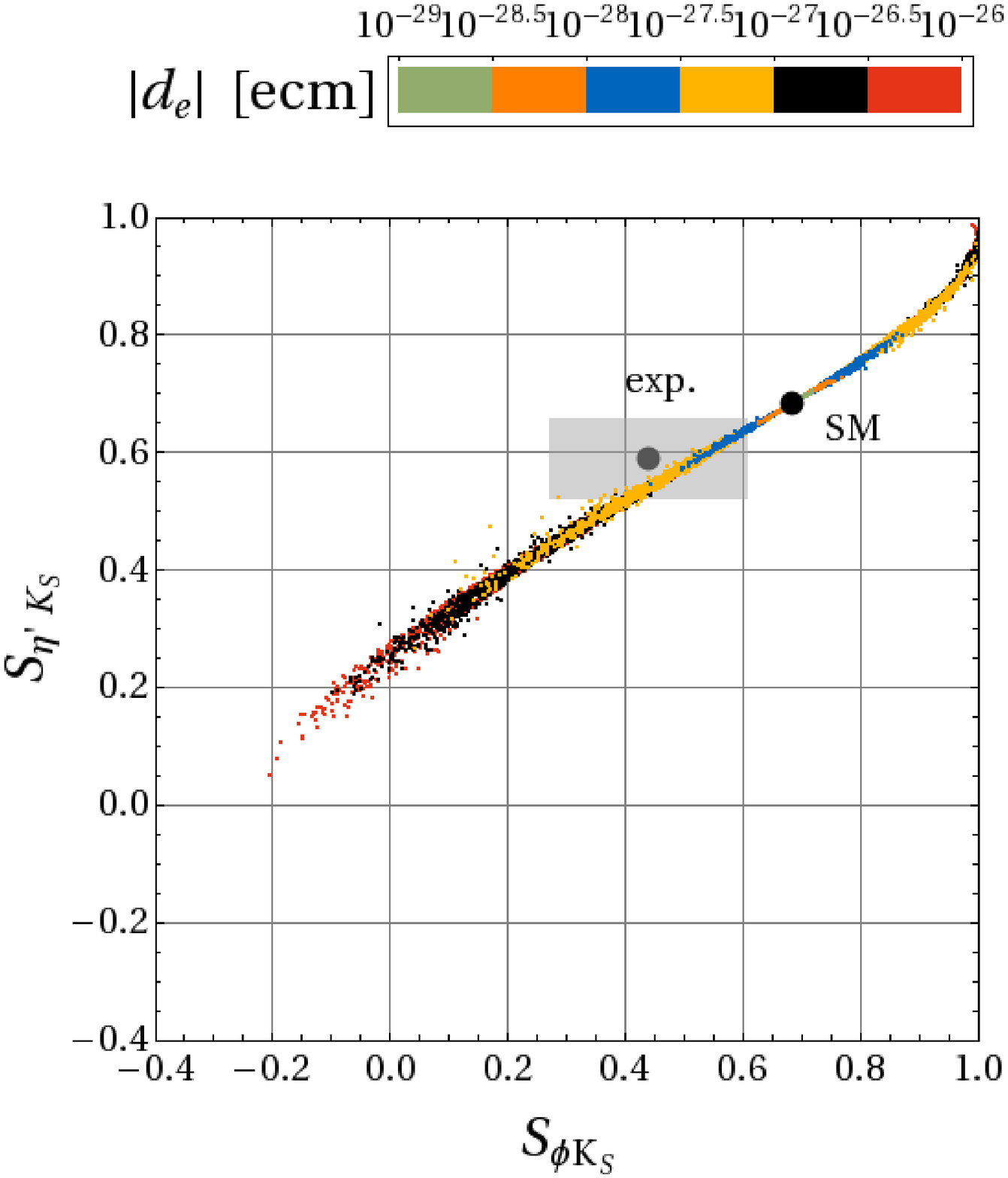} ~~~~
\includegraphics[width=.295\textwidth]{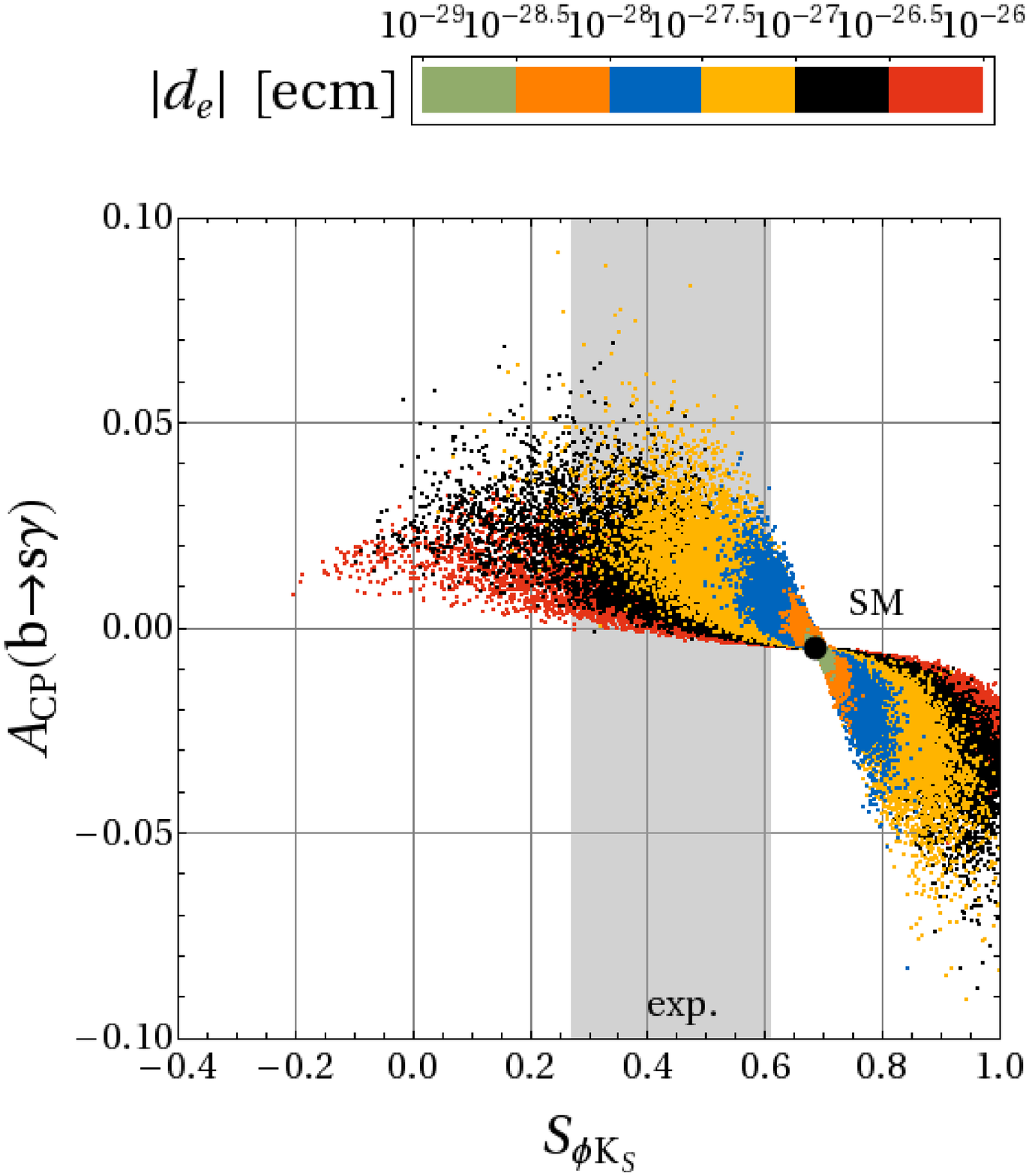} ~~~~
\includegraphics[width=.305\textwidth]{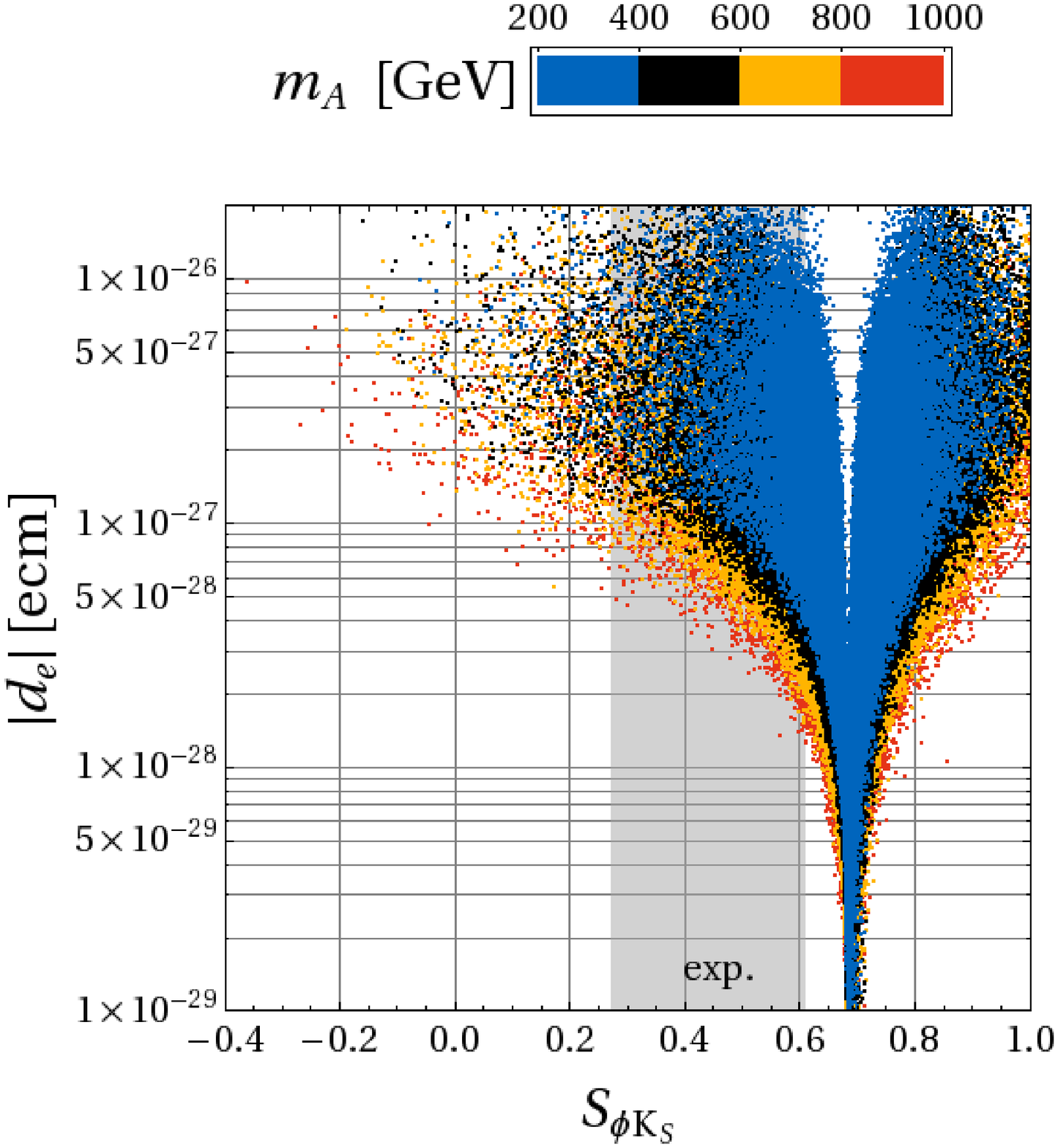}
\caption{Correlations of $S_{\eta^\prime K_S}$ with $S_{\phi K_S}$ (left), $A_{CP}(b\to s\gamma)$ with $S_{\phi K_S}$ (middle) and $d_e$ with $S_{\phi K_S}$ (right) in the FBMSSM. The gray regions correspond to the experimental $1\sigma$ ranges for $S_{\phi K_S}$ and $S_{\eta^\prime K_S}$. 
}
\label{fig:FBMSSM}
\end{figure}

We find that in this framework non-standard effects in CP violating $b\to s$ observables arise dominantly through the $\Delta F=1$ magnetic and chromomagnetic dipole operators. The corresponding Wilson coefficients get NP contributions mainly through Higgsino - stop loops that are $\tan\beta$ enhanced and proportional to a single complex parameter combination $C_{7,8} \propto \mu A_t$. In fig.~\ref{fig:FBMSSM} we show three examples of the resulting highly correlated effects in low energy observables.

Both $S_{\phi K_S}$ and $S_{\eta^\prime K_S}$, that we evaluate following~\cite{Buchalla:2005us,Hofer:2009xb}, can depart significantly from their SM expectations. The effects in these CP asymmetries are strongly correlated and both observables can be brought simultaneously in agreement with the measurements.

The direct CP asymmetry in $b \to s \gamma$~\cite{Soares:1991te} is a very suitable observable to look for NP effects~\cite{Kagan:1998bh}, as it is predicted to be very small in the SM, $A_{CP}(b\to s\gamma) \simeq -0.4 \%$~\cite{Hurth:2003dk}. In the FBMSSM values up to $\pm 6\%$ can be reached. Furthermore, $S_{\phi K_S}$ in agreement with the central experimental value unambiguously implies a positive value for $A_{CP}(b\to s\gamma)$.

The NP effects in $S_{\phi K_S}$ are also strongly correlated with the EDMs of the electron and neutron. In our framework, the dominant contributions to $d_e$ and $d_n$ arise from two loop Barr-Zee type diagrams~\cite{Chang:1998uc} that are also proportional to Im$(\mu A_t)$. The desire to explain the measured value of $S_{\phi K_S}$ then implies lower bounds of $d_e$ and $d_n$ at the level of $10^{-28}e\,$cm, only one order of magnitude below the current experimental constraints.

Finally we also find large effects in several observables accessible in the $B \to K^* \mu^+ \mu^-$ decay~\cite{Bobeth:2008ij,ABBBSW}. In particular, as discussed in \cite{ABBBSW}, the two T-odd CP asymmetries $\langle A_7 \rangle$ and $\langle A_8 \rangle$ are strongly enhanced with respect to their tiny SM predictions. The effects in $\langle A_7 \rangle$ and $\langle A_8 \rangle$ are highly correlated among themselves and also with the other CP violating observables discussed so far.

Concerning CP violation in $\Delta F=2$ on the other hand, the leading contributions to the mixing amplitudes for the $K$, $B_d$ and $B_s$ systems are not sensitive to the new phases of the FBMSSM. In fact, only for an extremely light SUSY spectrum with Higgsinos and stops lighter than 200 GeV, $\epsilon_K$ can be modified by a positive NP shift at the level of at most $15\%$. Also $S_{\psi K_S}$ and $S_{\psi\phi}$ remain essentially SM like, with $0.03 \lesssim S_{\psi\phi} \lesssim 0.05$.

To summarize, we stress that the combined study of the above considered observables and especially the characteristic pattern of correlations among them constitutes a very powerfull test of the FBMSSM framework. In particular, if a large $B_s$ mixing phase will be confirmed at LHCb, the FBMSSM can eventually be ruled out.

\subsection{Introducing New Sources of Flavor Violation}

In order to generate sizeable effects in $S_{\psi\phi}$ in the MSSM, one has to go beyond the minimal ansatz of the FBMSSM and introduce not only additional sources of CP violation, but also of flavor violation. The latter can be present both in the soft masses of the squarks and in the trilinear couplings and they are conveniently parameterized by so called Mass Insertions (MIs). In presence of complex MIs, flavor and CP violating gluino-squark-quark interactions arise, that typically give the dominant contributions to FCNCs. 
While left-right flipping MIs are strongly constrained by the $b \to s\gamma$ decay and can hardly generate effects in $B_s$ mixing, $S_{\psi\phi}$ can take values in the entire range $-1 < S_{\psi\phi} < 1$ if left-left and/or right-right MIs are present. In particular, if both left-left and right-right MIs are present simultaneously, contributions to the $B_s$ mixing amplitude are generated that are strongly enhanced by renormalization group effects~\cite{Ciuchini:1997bw} and a large loop function~\cite{Gabbiani:1996hi}. In such a situation, even for moderate, CKM like values of the MIs $\sim |V_{ts}|$, huge effects in $S_{\psi\phi}$ can be achieved.

\section{Summary and Outlook}

In a flavor blind MSSM, sizeable non-standard effects in CP violating low energy observables are possible. In particular CP violating $\Delta F=0$ and $\Delta F=1$ dipole amplitudes can receive large complex NP contributions, leading to highly correlated modifications of the SM predictions of the EDMs, $S_{\phi K_S}$, $S_{\eta^\prime K_S}$, $A_{CP}(b\to s\gamma)$ and CP asymmetries in $B \to K^* \mu^+ \mu^-$. CP violation in $\Delta F=2$ amplitudes however remains SM like, {\it i.e.} one gets only small effects in $\epsilon_K$, $S_{\psi K_S}$ and especially in $S_{\psi\phi}$.

To generate large CP violating effects in $\Delta F=2$ amplitudes, additional flavor structures in the soft SUSY breaking terms are required. As left-left MIs are always induced radiatively through renormalization group running, models predicting sizeable right-right MIs are natural frameworks where a large $B_s$ mixing phase can occur. A detailed comparative study of the phenomenology of well motivated SUSY flavor models showing representative patterns of mass insertions can be found in~\cite{ABGPS}.

\begin{theacknowledgments}

I warmly thank the other authors of~\cite{ABP},~\cite{ABBBSW} and~\cite{ABGPS} for the very pleasant collaborations. This work has been supported by the German Bundesministerium f\"ur Bildung und Forschung under contract 05HT6WOA and the Graduiertenkolleg GRK 1054 of DFG.

\end{theacknowledgments}



\end{document}